\documentclass[%
reprint,superscriptaddress, amsmath,amssymb,aps,prx,linenumbers,
floatfix,]{revtex4-2}
\usepackage{graphicx}
\usepackage{dcolumn}
\usepackage{bm}
\usepackage{hyperref}
\hypersetup{colorlinks=true, citecolor=blue, urlcolor=blue, linkcolor=blue}
\usepackage{braket}
\usepackage{accents}
\usepackage{multirow,microtype}
\usepackage{dcolumn,booktabs}
\newcolumntype{d}[1]{D{.}{.}{#1}}
\usepackage[normalem]{ulem}
\usepackage{xcolor}
\definecolor{mycol}{RGB}{0,0,160} 
\usepackage{float}
\usepackage{bbold}
\usepackage[utf8]{inputenc}

\UseRawInputEncoding

\usepackage{mathtools}

\usepackage{academicons}
\usepackage{xcolor}

\usepackage{academicons}
\definecolor{orcidlogocol}{HTML}{A6CE39}

\usepackage{orcidlink}

\begin{document}

\title{Learning the Intrinsic Dimensionality of Fermi-Pasta-Ulam-Tsingou Trajectories: A Nonlinear Approach using a Deep Autoencoder Model} 

\author{Gionni Marchetti\,\orcidlink{0000-0002-8045-5878}}

\affiliation{Barcelona, Spain}

\email{gionnimarchetti@gmail.com}

\date{\today}
\nolinenumbers

\begin{abstract}
We address the intrinsic dimensionality (ID) of high-dimensional trajectories, comprising $n_s = 4\,000\,000$ data points, generated by the Fermi-Pasta-Ulam-Tsingou (FPUT) $\beta$ model with $N = 32$ oscillators. To this end, a deep autoencoder (DAE) is employed to infer the ID in the weakly nonlinear regime, where energy recurrences are observed ($\beta \lesssim 1$). Our results indicate that the trajectories lie on a two-dimensional Riemannian manifold ($m^{\ast}=2$) embedded in a $64$-dimensional phase space. This finding is independently corroborated by a Fourier analysis of the time series associated with the normal mode coordinates and their corresponding energies, which reveals two independent  frequencies, consistent with quasi-periodic motion on a two-dimensional invariant torus $\mathbb{T}^2$. By contrast, principal component analysis (PCA) can  provide only a reasonable upper bound on the intrinsic dimensionality. Furthermore, the DAE predicts that the intrinsic dimensionality increases to $m^{\ast}=3$ at $\beta = 1.1$, coinciding with the onset of a symmetry-breaking (SB) phenomenon characteristic of the $\beta$ model, in which additional modes with even wave numbers ($k=2,4$) become excited. Notably, this symmetry-breaking transition is not detected by the linear PCA-based approach.
\end{abstract}

\maketitle 

\section*{}
\textbf{The Fermi-Pasta-Ulam-Tsingou (FPUT) model was introduced to investigate the expected ergodic behaviour of a one-dimensional chain of nonlinearly coupled oscillators. However, the energy recurrences first observed in computer simulations in 1953 appear to contradict the \emph{ergodic hypothesis}. In the present work, using a deep autoencoder (DAE), we show that, in the weakly nonlinear regime ($\beta \lesssim 1$), the high-dimensional trajectories of the FPUT $\beta$ model with $N = 32$ oscillators, consisting of $n_s = 4\,000\,000$ data points, lie on a Riemannian manifold with an estimated intrinsic dimensionality (ID) of $m^{\ast}=2$, embedded in the $64$-dimensional phase space. This estimate is independently supported by a Fourier analysis of the normal mode coordinates $Q_i$  ($i=1, 3, 5$) and their corresponding energies $E_i$, which reveals two independent  frequencies, consistent with quasi-periodic motion on a two-dimensional invariant torus $\mathbb{T}^2$. Furthermore, the DAE predicts an increase of the intrinsic dimensionality to $m^{\ast}=3$ when $\beta = 1.1$. We argue that this increase is associated with the onset of a symmetry-breaking (SB) phenomenon in the $\beta$ model, whereby additional modes with even wave numbers ($k=2,4$) begin to acquire a significant fraction of the system's energy.}

\section{Introduction}\label{sec:introduction} 

Fermi believed that weakly coupled nonlinear systems would inevitably exhibit ergodic behavior, as required to approach thermal equilibrium~\cite{Ford1963}. As a consequence, in 1953, the first computer simulations of the Fermi-Pasta-Ulam-Tsingou (FPUT) model, a one-dimensional chain of weakly nonlinearly coupled oscillators, were started aiming to study the system's thermalization, which in turn should have confirmed the validity of the \emph{ergodic hypothesis}~\cite{fermi1955,FORD1992,weissert1997, Gallavotti2008FPU, lepri2023, Dongarra2024}. In contrast, the energy recurrence phenomenon observed from simulations, for which the total energy was shared between a few energy modes only, seemed to contradict the ergodic hypothesis, a fact often referred to as the FPUT paradox.
 
Since then, the FPUT model has had a huge impact on our understanding of nonlinear dynamics and deterministic chaos. It would be an impossible task to mention all the important contributions made in the last seventy years attempting to explain the peculiar dynamical features of the FPUT chain. Here, we limit ourselves to recalling that the recurrence phenomenon can be linked to the Kolmogorov--Arnol'd--Moser (KAM) theorem~\cite{Kolmogorov1954, Arnold1963, Moser1962}, the Toda lattice~\cite{Toda1967, Benettin2013}, the solitonic behaviour of the Korteweg--de Vries (KdV) equation~\cite{Zabusky1965}, and q-breathers~\cite{Flach2005, Flach2006}. Additionally, there exist various theories and methods attempting to explain how the system eventually approaches equilibrium: the existence of a stochastic threshold~\cite{Izrailev1966, Lvov2018}, the FPUT metastability~\cite{fucito1982, Gallavotti2008FPU, reiss2023}, and the spectral-entropy equipartition indicator~\cite{Livi1985, Gallavotti2008FPU}.

Recently, a data-driven approach based on principal component analysis (PCA) was proposed to infer the intrinsic (or effective) dimensionality (ID) $m^{\ast}$ of the high-dimensional trajectories of the FPUT $\beta$ model with $N=32$ oscillators~\cite{marchetti2025}. This work was chiefly motivated by the observation that the ergodic hypothesis requires trajectories to lie on a constant-energy hypersurface of dimension $n-1$ in $\mathbb{R}^n$ ($n=2N$)~\cite{huang1987}. As a result, thermal equilibrium cannot be reached when the trajectories lie on very low-dimensional manifolds, a scenario which aligns with the application of the KAM theorem to the FPUT model in a weakly nonlinear regime~\cite{BALDOVIN2025}. Indeed, according to the KAM theorem, the orbits should be constrained to low-dimensional invariant tori~\cite{Ford1970, RINK2003}. 

In this regard, PCA when applied to entire trajectory data consisting of $n_s = 4\,000\,000$ data points, with the initial condition corresponding to excitation of only the first energy mode ($k=1$), estimates the ID to be either $2$ or $3$ when $\beta \lesssim 1.1$~\cite{marchetti2025}. These findings strongly suggest non-ergodic behavior, and are consistent with other qualitative observations in the same regime that support the underlying quasi-periodic motion of the FPUT system: the hexagonal patterns shown by Poincar\'{e} maps~\cite{giordano2006}, and the energy recurrences observed from the simulations~\cite{marchetti2025}.

However, being linear, PCA cannot reliably estimate the intrinsic dimensionality of nonlinear data~\cite{Joliffe2016,hastie2017,Lever2017,Greenacre2022,meila2024, Shinn2023}. Indeed, it assumes that the data points lie on optimal hyperplanes of a given rank $m$ (or, equivalently, on $m$-dimensional linear subspaces if the data are centered). This assumption is not necessarily satisfied, as shown by a qualitative analysis of early-stage orbits performed using $t$-distributed stochastic neighbor embedding ($t$-SNE), a nonlinear manifold-learning algorithm~\cite{vandermaaten2008,Kobak2019,Kobak2021,debodt2025,marchetti2025}. More importantly, the inherent limitation of a linear approach was further confirmed by the analysis of trajectory data for $\beta = 0.1$ using multi-chart flows, a manifold-learning method~\cite{Yu2025}. Specifically, this state-of-the-art method suggests that the trajectory lies near a  Riemannian manifold of dimension $m^{\ast} = 2$.
 
With this in mind, it makes sense to replace PCA with its nonlinear generalization: the autoencoder (AE)~\cite{goodfellow2016,Ranzato2007}. The AE is a type of artificial neural network (ANN) capable of learning the nonlinear structure of data. Specifically, we employ a relatively simple undercomplete deep autoencoder (DAE) to learn the actual nonlinear structure of the trajectory data under scrutiny (see Sec.~\ref{sec:DAE} for details). We find that this model can learn and generalise appropriately, yielding negligible reconstruction errors in comparison to those obtained using PCA. From the DAE reconstruction error curves, the corresponding elbow points estimate $m^{\ast} = 2$ for all available trajectory data associated with the recurrence phenomenon ($\beta \lesssim 1$) (see Fig.~\ref{fig:fig5}). Such a finding is independently supported by a Fourier analysis of the normal mode coordinates $Q_i$ and their corresponding energies $E_i$ ($i=1,3,5$) (see Sec.~\ref{sec: spectra} for details). The identification of two independent frequencies is consistent with the system performing quasi-periodic motion on a two-dimensional invariant torus $\mathbb{T}^2$.

 Finally, the DAE predicts that the intrinsic dimensionality increases by one, i.e., $m^{\ast} = 3$, when $\beta = 1.1$. This change is associated with the onset of a symmetry-breaking (SB) phenomenon characterizing the $\beta$ model, in which additional energy modes with even wave numbers $k = 2, 4$ begin to share the total energy of the system. By contrast, this significant variation of ID is completely missed by PCA, for which the estimated dimensionality remains constant as $\beta \to 1.1$.

 \section{FPUT $\beta$ Model}\label{sec:fput_model}

\begin{figure}
   \includegraphics[width=\columnwidth]{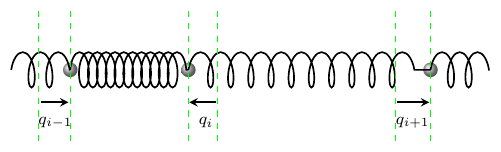}
\caption{The FPUT chain consists in a discrete system of equal mass points connected by springs. The coordinate $q_i$ represents the displacement of $i$-th point from equilibrium. }
\label{fig:fig1}      
\end{figure}

The FPUT  model describes a one-dimensional chain of $N$ nonlinear oscillators whose spring constants and masses are set to unity (see Fig.~\ref{fig:fig1}).
The $\beta$ ($\beta > 0$) model has the following Hamiltonian $H\left(q, p\right)$ where $q=\left(q_1, \cdots, q_N\right)$ and $p=\left(p_1, \cdots, p_N\right)$~\cite{BENETTIN2011}

\begin{equation}\label{eq:hamiltonian}
\begin{aligned}
H\left(q, p\right) ={} & \frac{1}{2} \sum_{i=1}^{N} p_{i}^{2} + \frac{1}{2} \sum_{i=0}^{N} \left(q_{i+1} - q_i\right)^{2}  \\
      & + \frac{\beta}{4} \sum_{i=0}^{N}  \left(q_{i+1} - q_i\right)^{4} \, .
\end{aligned}
\end{equation}

Accordingly, the equations of motion of the $\beta$ model are the following~\footnote{The equations of motion are commonly referred to as the canonical Hamilton equations.}

\begin{equation}\label{eq:firstOrder}
\begin{dcases}
\dot{q}_i = p_i, \\[1ex]
\dot{p}_i = q_{i+1} + q_{i-1} - 2q_i
  + \beta\Big[(q_{i+1}-q_i)^3 - (q_i-q_{i-1})^3\Big].
\end{dcases}
\end{equation}

Note that in Eq.~\ref{eq:firstOrder} the index $i$ runs from $0$ to $N +1$. This is why we assume that the chain has fixed ends, i.e., $q_0=q_{N+1} = 0$, and hence $p_0=p_{N+1} = 0$ (Dirichlet boundary conditions).  

From a mathematical point of view, the Cauchy problem for the first-order system in Eq.~\ref{eq:firstOrder} is well posed; that is, the solution exists and is unique because the vector field is analytic, being polynomial, since the number of oscillators $N$ is finite~\cite{Maspero2025}. However, the solution can be obtained only through numerical integration.
To this end, we used the velocity St\"{o}rmer-Verlet scheme, which is a second-order symplectic integrator well suited for Hamiltonian systems such as FPUT chains~\cite{Hairer2006, BENETTIN2011}. Consequently, in the present work, the trajectories were generated with high accuracy using a fixed time step of 0.05~\cite{giordano2006}.

Finally, although the analysis presented in this work refers only to the parameter $\beta$, the effective nonlinearity of the FPUT $\beta$ model is determined by the product $\beta\epsilon$, where $\epsilon$ denotes the energy density~\cite{marchetti2025}. Since the energy density is kept fixed throughout this study, the nonlinear strength of the system is uniquely controlled by $\beta$. Therefore, in the following, we report only the values of $\beta$.

\section{A Deep Autoencoder Model}\label{sec:DAE}

\begin{figure}
   \includegraphics[width=\columnwidth]{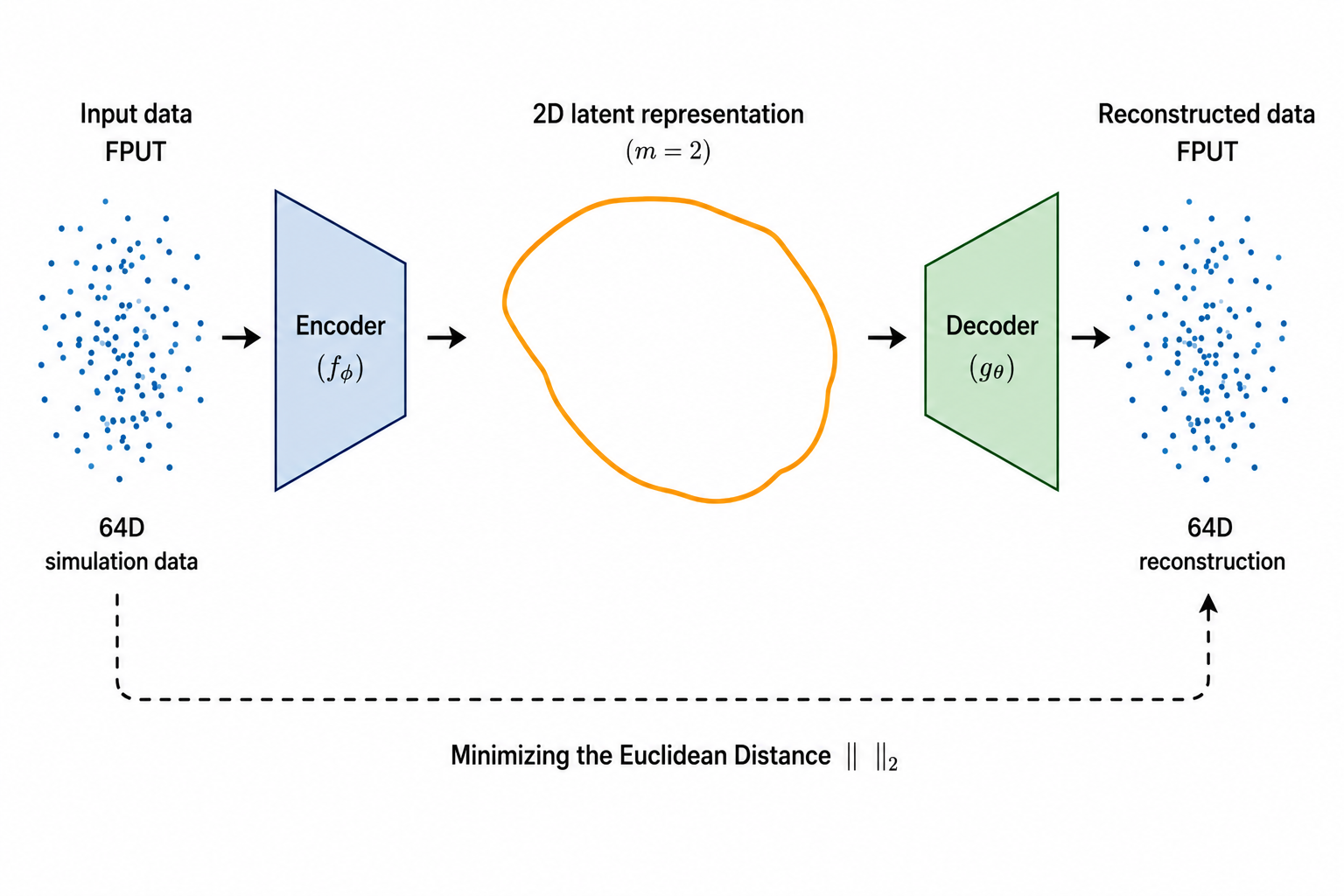}
\caption{Schematic view of a general autoencoder with bottleneck layer of dimension $m=2$ applied to FPUT simulation input data of dimension $n=64$. Note that  closed orbit-like curve (the embedding) in the two-dimensional latent space reproduces the one observed for an early-stage trajectory consisting of $10, 000$ data points, assuming $\beta =0.1$.  }
\label{fig:fig2}      
\end{figure}

A general autoencoder is illustrated schematically in Fig.~\ref{fig:fig2}. Each high-dimensional input, here the $64$-dimensional FPUT simulation data $\mathbf{x}\in\mathbb{R}^{64}$, is first mapped by the encoder, a neural network denoted by the mapping $f_\varphi$ and parametrized by the weights $\varphi$, to a low-dimensional latent representation $\mathbf{z}=f_\varphi(\mathbf{x})\in\mathbb{R}^{m}$, where $m$ denotes the bottleneck dimension. This bottleneck forces the network to learn the most informative coordinates describing the underlying data manifold, depicted as the closed orbit-like curve in the latent space with $m=2$, consistent with the quasi-periodic dynamics of the FPUT model in recurrence regime, i.e., when $\beta \lesssim 1$. The decoder, another neural network denoted by the mapping $g_\theta$ and parametrized by the weights $\theta$, subsequently maps the latent representation back to the original $64$-dimensional space, producing its reconstruction $\hat{\mathbf{x}}=g_\theta(\mathbf{z})$. The encoder and decoder are trained jointly by optimizing the parameters $\varphi$ and $\theta$ to minimize the Euclidean reconstruction error $J_m$ (see Eq.~\ref{eq:rec_error}), or equivalently the mean squared error (MSE) loss, between the inputs and their respective reconstructions. In this way, the autoencoder learns an encoder-decoder pair that preserves the essential information required to reconstruct the data while passing through the $m$-dimensional bottleneck.

By training a family of autoencoders with different bottleneck dimensions $m$ and monitoring the reconstruction error $J_m$ as a function of $m$, the intrinsic dimensionality $m^{\ast}$ is identified as the point beyond which the error curve saturates, indicating that increasing the latent dimension no longer leads to a significant improvement in reconstruction accuracy~\cite{tenenbaum2000}. Consequently, $m^{\ast}$ is interpreted as the intrinsic dimensionality of the underlying data manifold. It is worth noting that AEs can be regarded as nonlinear generalizations of PCA. Indeed, it can be shown that a linear autoencoder is mathematically equivalent to PCA~\cite{Lindholm2022, Refinetti2022}.

In the present work, we shall employ a deep autoencoder model with five hidden layers $N_h = 5$, whose architecture with layer widths $[64, 32, 16, m, 16, 32, 64]$. Note that its architecture consists of two sequential networks: the encoder and the decoder. The first maps  $\mathbf{x}$ to  $\mathbf{z}$, while the second maps  $\mathbf{z}$ to $\mathbf{\tilde{x}}$.

In the following, we describe all the model hyperparameters, which were chosen based on their excellent reconstruction performance of the input data for latent dimensions $m \geq 2$, together with a preference for simpler models (Ockham's razor)~\cite{Jefferys1992, Ghahramani2015}. 

The number of neurons (or nodes) per layer, together with the respective activation functions are shown in Table~\ref{tbl:dae}. Note that all hidden neurons use  ReLU (Rectified Linear Unit) activation function, i.e., $h(y) = \max(0, y)$, for learning nonlinear mappings~\cite{Nair2010}. By contrast, the output layer employs a linear activation function, which is commonly used for regression tasks.

\begin{table}[ht]
    \begin{center}
        \begin{tabular}{ccc}
            \hline
	        Layer Type & Number of Neurons & Activation Function \\\hline
            input layer & 64 & - \\\hline
            encoder layer 1 & 32 & ReLU \\\hline        
            encoder layer 2 & 16 & ReLU \\\hline        
            bottleneck layer & $m$ & ReLU \\\hline        
            decoder layer 1 & 16 & ReLU \\\hline        
            decoder layer 2 & 32 & ReLU \\\hline        
            output layer & 64 & Linear \\\hline        
        \end{tabular}      
    \end{center}
    \caption{Number of neurons and corresponding activation functions of the DAE architecture under consideration. The size of the bottleneck layer, $m$, is assumed to satisfy $1 \leq m \ll n$, with $n = 64$.}
    \label{tbl:dae}
\end{table}

The model is trained using the Adam optimizer~\cite{Kingma2015} with a learning rate of $0.001$ and a mini-batch size of $256$, noting that smaller batch sizes (e.g., $32$) do not produce significant differences in performance. The number of training iterations is set to $300$ epochs. However, early stopping (ES), which acts as an implicit regularizer with a patience of $30$ epochs, is chiefly used to avoid unnecessary training, thereby saving significant time and computational resources. 
Additionally, the ES will retain the model corresponding to the minimum validation error during the optimization of the trainable parameters (weights and biases)~\footnote{Note that when the model's bottleneck has a single neuron, i.e., $m=1$, there are $4, 801$ trainable parameters.}. We implemented this model  using Keras and TensorFlow~\cite{Geron2017,chollet2018,abadi20216}.

In general, especially for neural networks, the training reconstruction error can be very low because the model may fit the training data extremely closely, thereby leading to serious overfitting. As a result, the model may not perform well on unseen data. Accordingly, to ensure that the model generalizes successfully, we split each trajectory dataset into a training set and a test set~\cite{Hauberg2025}. The training set contains $70\%$  of the input data, consisting of data points chosen at random, while the test set contains the remaining $30\%$  of the data. Note that $20\%$ of the training set is also used as validation set for monitoring the model's generalization during training. With this choice, we find that the difference between the training and test reconstruction errors is generally much less $1\%$.

To estimate ID from the reconstruction error curves~\cite{Cattell1966}, we need to restrict the latent dimensionality to as low as a single variable
($m=1$). In such a case, it would be unrealistic to expect an autoencoder to capture the complex structure of the
$64$-dimensional input data, unless the data lie on an approximately one-dimensional manifold~\cite{Lindholm2022}, which is not necessarily the case since we estimated ID to be greater than unity ($m^{\ast} = 2, 3$). Accordingly, this circumstance will cause a severe information-bottleneck regime, in which the DAE model is expected to perform poorly, but still significantly better than PCA. We note that 
this limitation is inherent to the constraint $m=1$ itself, and holds independently of the specific autoencoder architecture employed as shown in next Sec.~\ref{sec:discussion}. Nevertheless, the aforementioned limitation disappears altogether when the latent dimension $m \geq 2$.

We conclude noting that there is considerable room for improvement over the present model, for example by employing learning-rate decay, alternative activation functions such as GeLU, Mish, or SeLU, or a deeper network architecture~\cite{Fabio2026}. To illustrate this, a comparison with two different deeper models with $N_h=7$ is briefly addressed in the next section (Sec.~\ref{sec:discussion}). Nevertheless, our preference for the simpler model (ReLU, $N_h=5$, Adam) is justified by the reasonably accurate results achieved at a relatively low computational cost.

\subsection{Datasets and Data Preprocessing}\label{sec:datasets} 

Each dataset of size $n_s = 4\,000\,000$ corresponds to a set $\mathcal{X} = \{\mathbf{x}_1, \mathbf{x}_2, \cdots, \mathbf{x}_i, \cdots, \mathbf{x}_{n_s}\}$, whose elements $\mathbf{x}_i \in \mathbb{R}^n$ are data points that belong to a single trajectory obtained from numerical simulations of the FPUT system at a given value of $\beta \in [0.1, 1.1]$, sampled at intervals of $\Delta \beta =0.1$~\cite{marchetti2025}.

The set $\mathcal{X}$ is then conveniently arranged in a $n_s \times n$ data matrix $X$. However, before feeding it into the DAE model, a data preprocessing step is required, consisting of centering and scaling each variable by its standard deviation, as is customary in PCA~\cite{Joliffe2016}. First, centering the data, corresponding to setting the mean of each variable to zero, allows the PCA eigenvalues to be efficiently computed via singular value decomposition (SVD) (see Sec.~\ref{sec:pca}). Second, rescaling ensures that each variable, now having unit variance, contributes equally to the analysis regardless of its original scale~\cite{Joliffe2016}. This preprocessing step allows for a direct comparison between the DAE results and those obtained from the principal component analysis in Ref.~\cite{marchetti2025}.

\section{Results and Discussion}\label{sec:discussion}

We briefly recall the rationale motivating this work. According to Hamiltonian dynamics, the trajectories of the FPUT model evolve on the constant-energy hypersurface $\Sigma_E$, whose intrinsic dimensionality is $n-1$, where $n$ denotes the dimension of the phase space. However, the observed energy recurrences are consistent with quasi-periodic motion on an invariant torus $\mathbb{T}^{\tilde{m}}=S^1\times S^1\times\cdots\times S^1$, where $S^1$ denotes the unit circle and $\tilde{m}$ is the number of independent circle factors, equivalently the dimension of the torus. Such a torus is naturally parametrized by the angular coordinates $\varphi_1,\varphi_2,\ldots,\varphi_{\tilde{m}}$. By contrast, the DAE learns a latent parametrization of this nonlinear manifold through the variables $z_1,z_2,\ldots,z_{m^{\ast}}$, where the optimal latent dimensionality $m^{\ast}$ is determined from the reconstruction error using the elbow method. In the following, we show that the DAE estimate coincides with the dimension of the invariant torus, obtained from the Fourier analysis applied to normal mode coordinates $Q_i$ ($i=1, 3, 5$) and their corresponding energies $E_i$. In fact, we found that $m^{\ast}=\tilde{m}=2$.

Next, we compare the performance of two different DAEs with our model (ReLU, $N_h=5$, Adam), described in Section~\ref{sec:DAE}, to demonstrate that this choice is reasonable. Second, we focus specifically on the performance of our model. Accordingly, we show the learning curves obtained during model training on a given  dataset corresponding to $\beta = 0.3$. However, our analysis exemplifies the typical behaviour of the model across all the datasets under consideration.

 To begin, we compare our relatively simple model with two others featuring a deeper network structure ($N_h = 7$), with layer widths given by $[64, 48, 32, 16, m, 16, 32, 48, 64]$ for bottleneck sizes $m = 1$ and $m = 2$. The first model (ReLU, $N_h = 7$, Adam) differs only in the value of $N_h$, while the second introduces additional features: the GeLU (Gaussian Error Linear Unit) activation function and the SOAP (Simultaneous Orthogonalization and Preconditioning) optimizer with an exponential learning rate, replacing ReLU and Adam, respectively. The SOAP optimizer is a recently proposed method that accounts for correlations between parameters, whereas Adam treats them as independent~\cite{vyas2025}. The DAE (GeLU, $N_h = 7$, SOAP) incorporating SOAP, GeLU, and an exponential learning rate was suggested and implemented by Froheberg using the JAX library~\cite{jax2018github}\footnote{The learning rate follows an exponential decay schedule with an initial value of $10^{-3}$, decay rate $0.9$, and transition steps of $10^{5}$.}. 
 
\begin{figure}
    \centering
    \begin{minipage}{0.48\textwidth}
        \centering
        \includegraphics[width=\linewidth]{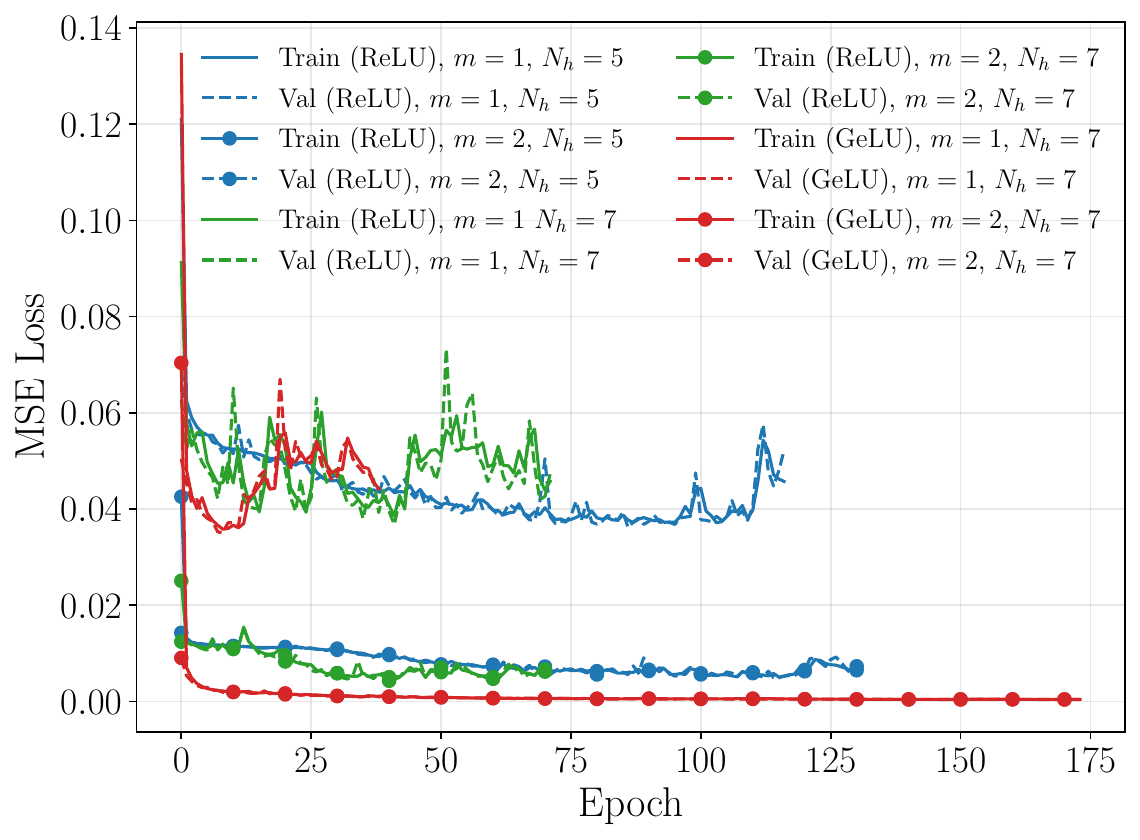}
        \caption{Training and validation MSE loss curves as functions of epoch for bottleneck sizes $m = 1, 2$ and trajectory data with $\beta = 0.3$ from DAEs: (ReLU, $N_h = 5$, Adam), (ReLU, $N_h = 7$, Adam), and (GeLU, $N_h = 7$, SOAP).  Throughout training, all deep autoencoders employ early stopping with a patience of $30$ epochs.}
        \label{fig:fig3}
    \end{minipage}
    \hfill
    \begin{minipage}{0.48\textwidth}
        \centering
        \includegraphics[width=\linewidth]{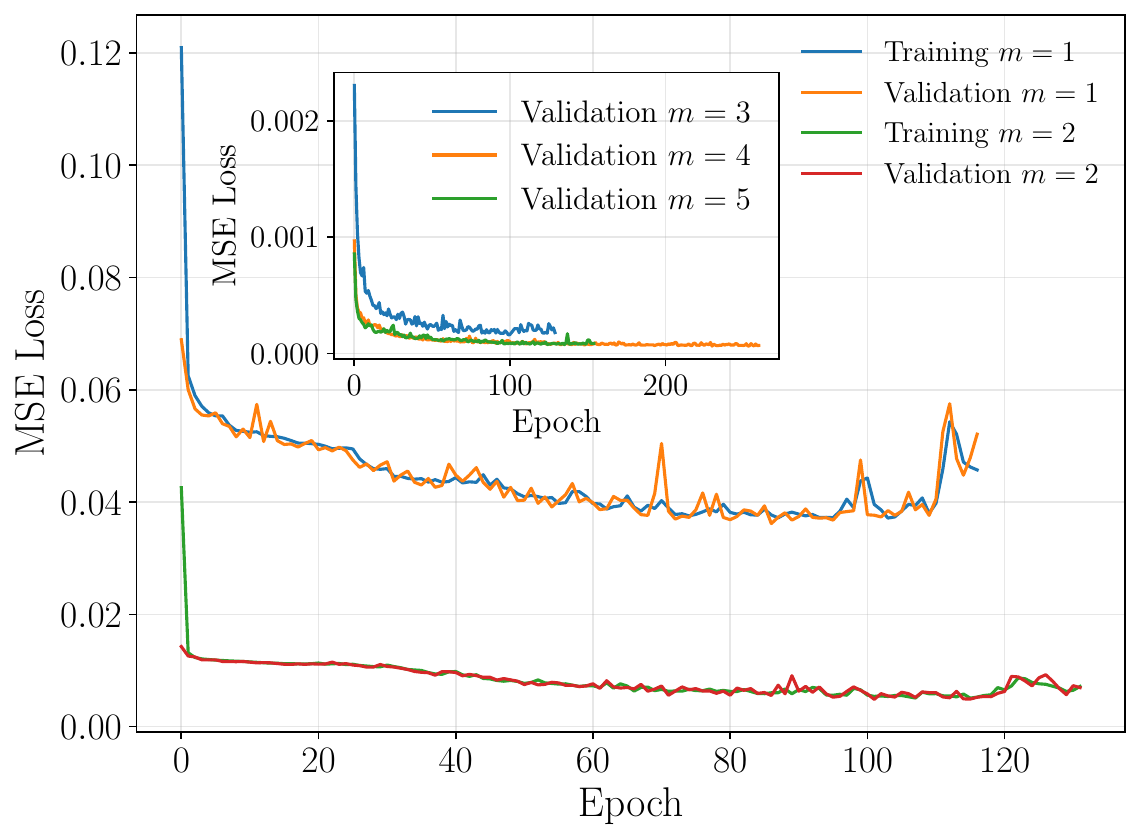}
        \caption{Training and validation MSE loss curves as functions of epoch for bottleneck sizes $m=1, 2$ and trajectory data with $\beta= 0.3$  from DAE (ReLU, $N_h = 5$, Adam). (Inset)
        Validation MSE loss curves as functions of epoch for $m=3, 4, 5$ and same data.  Throughout training, the deep autoencoder employs an early stopping with a patience of $30$.}
        \label{fig:fig4}
    \end{minipage}
\end{figure}

 In Fig.~\ref{fig:fig3} the training and validation MSE losses from the different DAEs are shown as functions of epoch for bottleneck sizes $m=1$ and $m=2$. Throughout, ES is set to $30$. The learning curves display significant qualitative and quantitative differences according to the latent dimension under scrutiny. Overall, we do not observe overfitting, as validation curves follow the training curves closely, yielding nearly identical errors. Accordingly, the models generalize well to unseen data.
 
However, when $m=1$, the irregular oscillations observed after a fast initial decay suggest that these autoencoders are approaching their capacity limit, signalling their inability to faithfully reconstruct the high-dimensional data. Note that regardless of the chosen DAE, the performance is nearly the same (MSE $\approx$ $0.036$--$0.038$). These facts should not be surprising considering that the ID will be shown to be $m^{\ast} =2$.

In contrast, when $m=2$, all curves display the expected learning trend: after a fast initial decay they plateau. In this case, the model with GeLU activation outperforms the other two, achieving a best MSE $\approx 4\times10^{-4}$, roughly one order of magnitude lower than the ReLU-based models ($\approx 4\text{--}5\times10^{-3}$).
However, the learning curves of our model (ReLU, $N_h = 5$, Adam) are smoother than those of the deeper model with $N_h = 7$, suggesting that increasing the network depth alone, without further changing other hyperparameters, does not guarantee better performance for the problem under scrutiny. Additionally, while all models errors are extremely low, especially when compared to those from PCA (see discussion below), the fact that the errors for $m=2$ can be as much as $80$ times smaller than those corresponding to $m=1$ strongly suggests that the intrinsic dimensionality exceeds one.
 
 Next, we turn to the training and validation MSE losses from our preferred model (ReLU, $N_h = 5$, Adam) which is significantly less expensive than the GeLU/SOAP model, the latter involving a deeper network and a more computationally demanding optimizer. The respective learning curves are shown as functions of epoch for bottleneck sizes $m=1, 2, \cdots, 5$ (see Fig.~\ref{fig:fig4}). In the main panel, it is now possible to better visualize these curves for when $m=1, 2$. Apart from the limitation of severe information-bottleneck regime which is evident in a noisy behaviour for $m=1$, this model generalizes well for both latent dimensions.
However, when $m = 1$, the loss saturates at a significantly higher value than for $m = 2$. Specifically, considering that early stopping is triggered at around epoch $116$, one finds
that the minimum validation MSE loss $\approx 0.036$ is attained at epoch $86$. This is roughly $7$ times larger than the corresponding best MSE loss observed when $m = 2$~\footnote{The best validation MSE loss is $\approx 0.0049$, attained at epoch $101$ when $m=2$.}.
By contrast, for $m \geq 2$ the model performs well, yielding substantially smaller reconstruction errors, as further supported by the well-converged validation loss curves for $m = 3, 4, 5$ shown in the inset of Fig.~\ref{fig:fig4}.

Next, the DAE reconstruction errors $J_m$  as a function of the latent dimension $m$ ($m=1, 2, \cdots$) for a given 
value of $\beta$, form a curve whose shape can help identifying the trajectory data's intrinsic dimensionality. Accordingly, the ID can then be inferred from the elbow point of such a curve~\cite{Cattell1966}. This heuristic is based on the assumption that, beyond the elbow, $J_m$ no longer decreases significantly as $m$ increases~\cite{tenenbaum2000}. This approach is inherently subject to a degree of ambiguity and subjectivity, but is still routinely used in the machine learning field, e.g., for manifold reduction and clustering~\cite{Geron2017}. In the following, we shall automate the identification of the elbow using the Kneedle algorithm (KA)~\cite{ville2011}, although in the present case the elbow points can be easily identified by visual inspection (which is not the case for strong nonlinearities; see discussion below).

\begin{figure}
   \includegraphics[width=\columnwidth]{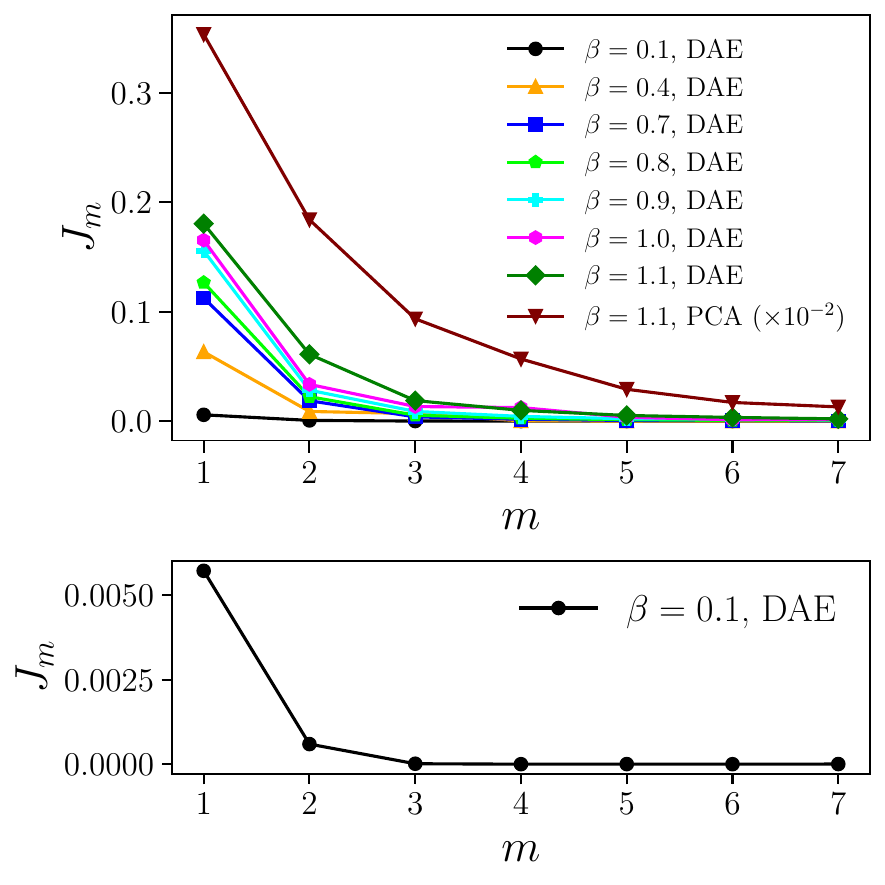}
\caption{(Top panel) The
DAE test reconstruction error curves as functions of the dimension $m$ for data corresponding to $\beta= 0.1, 0.4, 0.7, 0.8, 0.9, 1.0$, and $ 1.1$. The  PCA reconstruction error curve ($\beta= 1.1$) is rescaled by a factor of $100$ for proper visualization. (Bottom panel) The DAE  reconstruction error curve for data corresponding to $\beta= 0.1$.}   
\label{fig:fig5}      
\end{figure} 

\begin{figure*}
    \centering
    \begin{minipage}{0.48\textwidth}
        \centering
        \includegraphics[width=\linewidth]{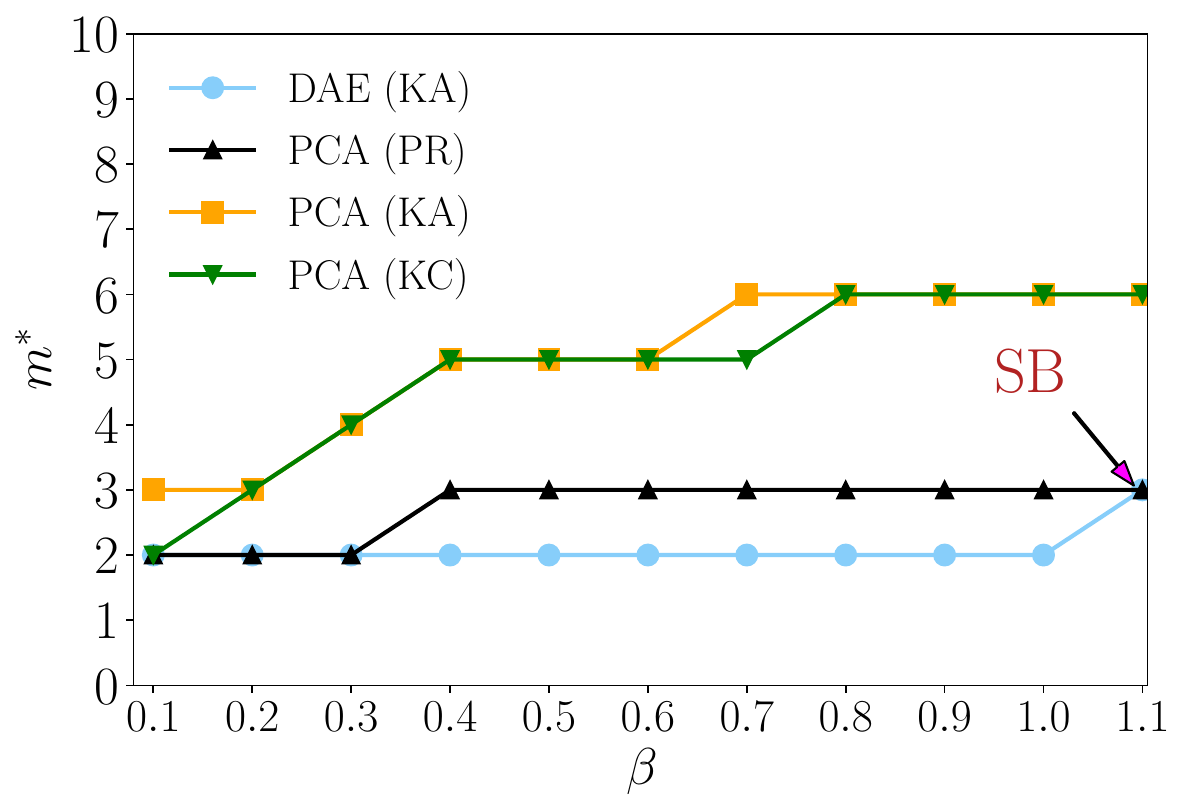}
        \caption{Estimated intrinsic dimension $m^{\ast}$ as function of $\beta \in [0.1, 1.1]$, using DAE model (ReLU, $N_h = 5$, Adam) with Kneedle algorithm (KA) and PCA with the following heuristics: Participation Ratio (PR) heuristic, Kneedle algorithm (KA), and Kaiser criterion (KC).}
        \label{fig:fig6}
    \end{minipage}
    \hfill
    \begin{minipage}{0.48\textwidth}
        \centering
        \includegraphics[width=\linewidth]{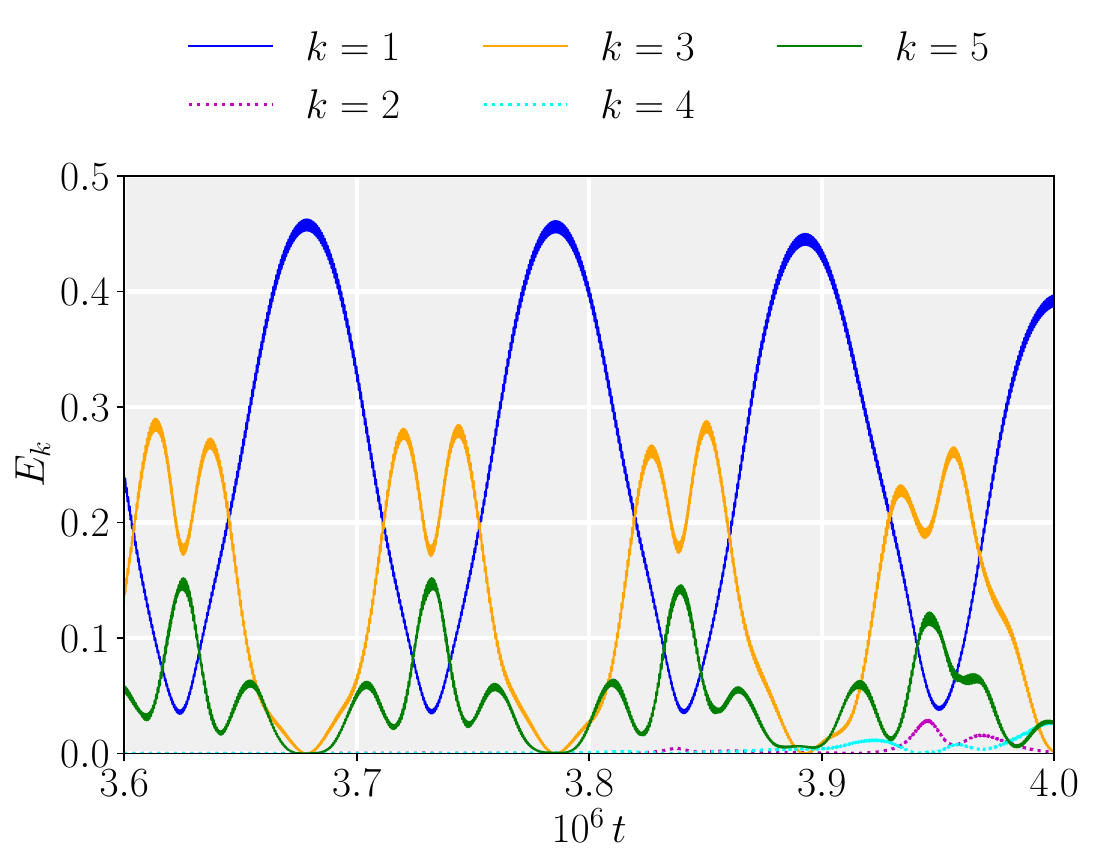}
        \caption{Time evolution of the energy $E_k$ of the Fourier modes with $k = 1, 2, 3, 4, 5$ for the $\beta$ model with $N = 32$ oscillators. The initial condition corresponds to assigning the energy $\mathcal{E}_1 \approx 0.45$ to the first mode $E_1$ only.}
        \label{fig:fig7}
    \end{minipage}
\end{figure*}

In Fig.~\ref{fig:fig5} (top panel), some selected representative reconstruction error curves as functions of $m$ ($m=1, 2, \cdots 7$) are shown. These curves correspond to $\beta = 0.1, 0.4, 0.7, 0.8, 0.9, 1$, and $\beta = 1.1$. Additionally, a curve from PCA reconstruction error corresponding to $\beta =1.1$ is displayed. The latter has been rescaled by a factor of $100$ to enable visualization alongside the DAE curves, confirming the superior reconstruction performance of the autoencoder. 
All DAE reconstruction error curves decrease monotonically, except for the one corresponding to $\beta = 0.1$, whose actual trend is not visible due to its small magnitude. For this reason, this curve is plotted separately in the bottom panel of Fig.~\ref{fig:fig5} to improve visualization. All curves for which $\beta \leq 1$ exhibit an evident elbow point at $m = 2$. Remarkably, this finding agrees with the approach based on multi-chart flows recently proposed by Yu et al.~\cite{Yu2025}. By contrast, when $\beta = 1.1$, the elbow point is shifted by unity. Therefore, our analysis suggests that $m^{\ast} = 2$ for $\beta \leq 1$, while $m^{\ast} = 3$ for $\beta = 1.1$.

We note in passing that one would expect a similar approach, based on elbow detection, to be applicable straightforwardly to reconstruction error curves at much larger values of $\beta$, not considered in the present work. Nevertheless, this is not the case, since as $\beta$ increases, no evident elbows emerge in the reconstruction error curve. This behaviour is analogous to that observed for PCA reconstruction error curves~\cite{marchetti2025}. To illustrate the above observations, the DAE and PCA curves corresponding to $\beta = 1.8$ are shown in  Fig.~\ref{fig:fig11} of Sec.~\ref{sec:TPUs}. 

That being said, alternative methods will be necessary to ensure a reliable estimation of the ID of trajectory data generated in the strongly nonlinear regime~\cite{Zeng_2024,Yu2025, zelong2025}.

We can now compare the intrinsic dimensionalities inferred from the DAE with those previously obtained using PCA~\cite{marchetti2025}. Since PCA admits several alternative heuristics to the Kneedle algorithm, we additionally consider the Kaiser criterion~(KC)~\cite{jolliffe2002pca} and the Participation Ratio~(PR) heuristic~\cite{Kramer1993}, both of which are computed directly from the PCA eigenvalues (see Secs.~\ref{sec:pca} and~\ref{sec:heuristics} for details on PCA and for the definitions of the heuristics, respectively).

In Fig.~\ref{fig:fig6}, we compare the estimated IDs obtained using the DAE model and KA (circles) with those from PCA together with PR (triangles), KA (squares), and KC (inverted triangles), respectively. First, we note that PCA combined with KA and KC substantially overestimates the intrinsic dimensionality for $\beta \geq 0.3$, since in these cases $4 \leq m^{\ast} \leq 6$. By contrast, PCA together with the PR heuristic yields $m^{\ast} = 2$ and $m^{\ast} = 3$ for $\beta \leq 0.3$ and $\beta > 0.3$, respectively. The latter values agree with those obtained using the autoencoder when $\beta \leq 0.3$ and $\beta = 1.1$, but overestimate the remaining cases by one.

We argue that this agreement is purely coincidental, as it is well known that PCA tends to overestimate the ID when the data have a nonlinear structure, as in the present case. A classical example is the synthetic Swiss roll dataset, for which PCA together with elbow detection predicts $m^{\ast} = 3$, whereas the data actually lie on a two-dimensional nonlinear manifold ($m^{\ast} = 2$)~\cite{Roweis2000, tenenbaum2000}. Based on general considerations, PCA is expected to overestimate IDs by at least one when the data are nonlinear~\cite{Zeng_2024}. Consequently, we conclude that, in the present case, the linear approach performs reasonably well only for $0.3 \leq \beta \leq 1$, as it yields a meaningful upper bound on the actual ID.

Next, we note that there is another important difference between the DAE approach and the linear one. Indeed, according to the latter, the estimated ID remains unchanged as $\beta \to 1.1$, irrespective of the heuristic used, whereas the DAE predicts an increase by unity, i.e., $m^{\ast} = 3$.  To understand this dimensionality variation, one needs to consider an important feature characterizing the $\beta$ model: the symmetry about its center for initially symmetric excitations~\cite{reiss2023, pace2019}. This means that if we initially excite a Fourier mode with an odd wave number, then the energy cannot spread to even-numbered modes. In the present case, the initial condition corresponds to assigning the energy $\mathcal{E}_1 \approx 0.45$ to the first mode $E_1$ only. As a result, the energy is expected to be shared only by the odd-numbered modes only, in the regime of recurrence. In our case, this occurs when $\beta \lesssim  1$, as illustrated for instance by the simulation for  $\beta = 1$ in Fig.~\ref{fig:fig9}. In contrast, when $\beta = 1.1$, the even-numbered modes $E_2$ and $E_4$ are also involved in the system's dynamics, as illustrated in Fig.~\ref{fig:fig7}. In such a case, the mode energies  $E_2$ and $E_4$, shown as dotted curves in purple and cyan, respectively, acquire a small amount of energy towards the end of the simulation. Note that this energy gain is not attributable to the symplectic integrator numerical accuracy. Indeed, in Sec.~\ref{sec:fput_facts} we show that  the energy drop of odd modes $E_1, E_3$ and $E_5$, which is approximately  $6.55\%$ of initial energy $\mathcal{E}_1$, quantitatively accounts for the energy transferred to even modes $E_2$ and $E_4$, corresponding approximately  $6.02\%$ of $\mathcal{E}_1$ (see Fig.~\ref{fig:fig10}). This energy transfer from odd to even modes occurs slowly in a regime where the system has not yet reached thermal equilibrium and cannot be attributed to the characteristic oscillatory time dependence of the symplectic integrator error~\cite{Donnelly2005}.

That being said, we can relate the change in ID from $m^{\ast} = 2$ to $m^{\ast} = 3$ to the occurrence of SB in the $\beta$ model, whereby energy begins to flow from odd to even modes.  Notably, this important dynamical feature could not be detected using PCA with standard heuristics.  We suspect that this change in ID due to symmetry-breaking may imply a substantial difference in the geometric or topological features of the 
$2$- and $3$-dimensional Riemannian manifolds on which the trajectories lie. Accordingly, such a difference could be properly addressed using suitable tools from topological and geometric data analysis~\cite{Carlsson2009, chazal2021, Otter2017, Munch_2017, Papillon2025, hickokthesis2023, hickok2023}.

\begin{figure}
\includegraphics[width=\columnwidth]{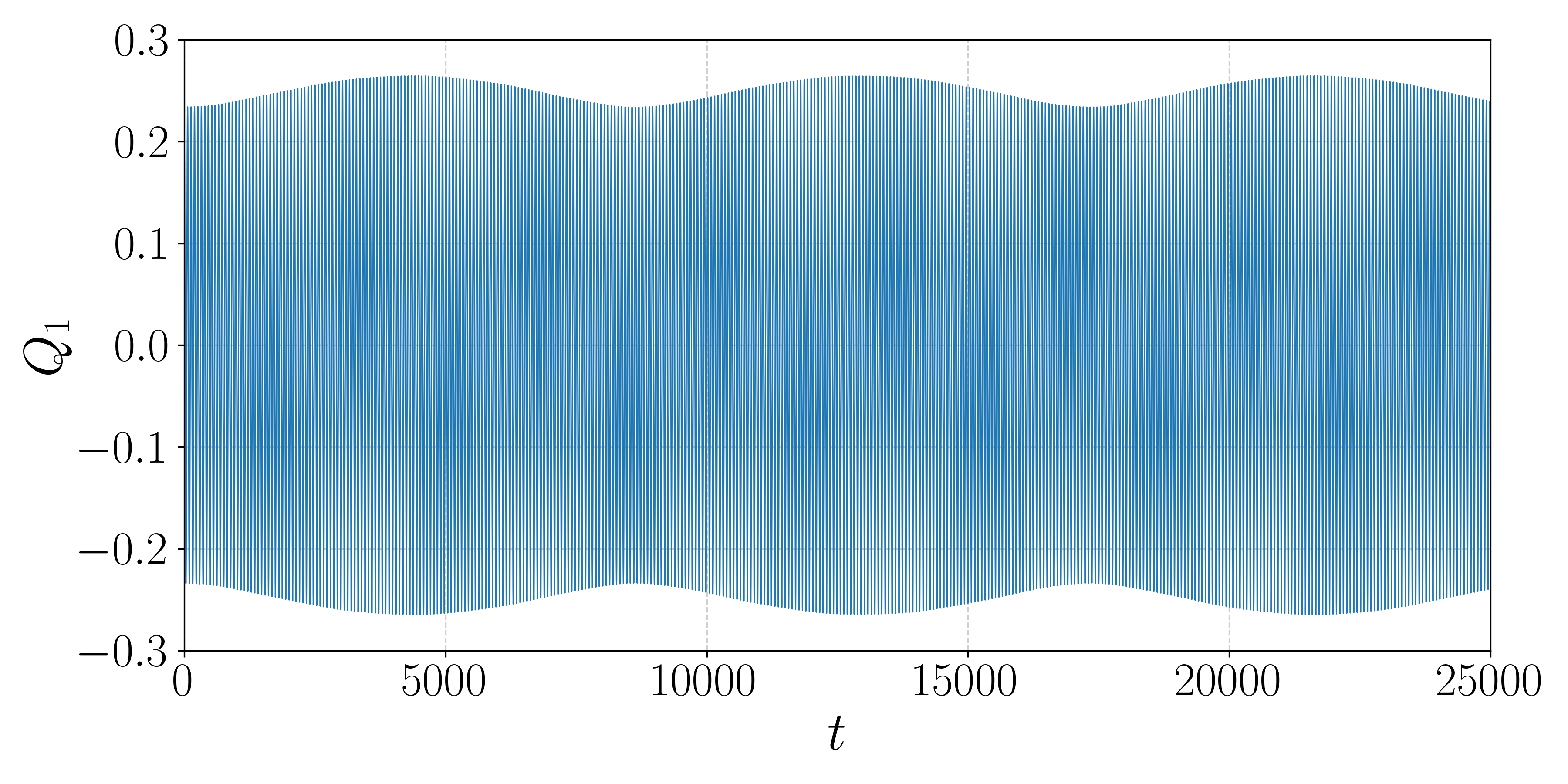}
\caption{ Time-evolution of the first normal mode $Q_1$ for FPUT chain ($N=32$), assuming $\beta=0.1$. The initial condition corresponds to assigning the energy $\mathcal{E}_1 \approx 0.45$  to the first mode $E_1$ only.}   
\label{fig:fig8}      
\end{figure}

In the recurrence regime ($\beta \lesssim 1$), our ID estimate, i.e., $m^{\ast}=2$, suggests that the nonlinear manifold on which the system evolves can be described by only two coordinates (or degrees of freedom). These correspond to the two latent variables, namely the codes $z_{1, \beta}$ and $z_{2, \beta} $, learned by the DAE during training. On the other hand, according to KAM theory, such a manifold is expected to be a two-dimensional invariant torus $\mathbb{T}^2$. Accordingly, only two angular coordinates, $\varphi_{1,\beta}$ and $\varphi_{2,\beta}$, are required to describe the free motion on $\mathbb{T}^2$, whose time evolution is governed by two independent frequencies associated with the FPUT Hamiltonian. In this regard, the time evolution of the normal mode coordinates $Q_1$, $Q_3$, and $Q_5$ exhibits beat patterns of different amplitudes (the beats corresponding to $Q_1$ for $\beta=0.1$ are shown in Fig.~\ref{fig:fig8}). These beats can be interpreted as arising from the superposition of at least two cosine waves with slightly different frequencies and amplitudes~\cite{Feynman1970Vol1}. Under this interpretation, the corresponding angular frequencies, denoted hereafter by $\omega_{1,\beta}$ and $\omega_{2,\beta}$, naturally correspond to the frequencies governing the time evolution of the angular coordinates $\varphi_{1,\beta}$ and $\varphi_{2,\beta}$, respectively~\footnote{Accordingly, $\varphi_{i,\beta}(t)=\varphi_{i,\beta}^{0}+\omega_{i,\beta}t$, where $\varphi_{i,\beta}^{0}$ ($i=1,2$) denotes the corresponding initial value.}

To determine the frequencies of interest, we performed a frequency analysis using the Fast Fourier Transform (FFT) (see Sec.~\ref{sec: spectra} for details). The analysis was applied to the datasets containing the time series of the normal mode coordinates $Q_1$, $Q_3$, and $Q_5$, together with their corresponding energies $E_1$, $E_3$, and $E_5$, obtained for $\beta=0.1$, that is, at the lowest nonlinearity considered in the present work.

The analysis reveals significant spectral peaks at the odd and even harmonics of the frequency $f_{1,\beta}\approx1.5170\times10^{-2}$ (arbitrary units) in the spectra of the signals $Q_i$ and $E_i$ ($i=1,3,5$), respectively (see the corresponding panels in Fig.~\ref{fig:fig13}). This frequency can be regarded as the first carrier frequency contributing to the observed beats. By contrast, the recurrence (beat) frequency, $f_{\rm rec,\beta}\approx1.15\times10^{-4}$ (arbitrary units), which modulates the beat amplitude, appears only as a very weak spectral peak, with an amplitude of approximately $0.002$ in the spectra of $E_1$ and $E_3$. Nevertheless, the same frequency can be inferred directly from the separation between successive maxima of the signals $Q_k$, or equivalently $E_k$, yielding the recurrence period $T_{\rm rec,\beta}\approx8680$ (arbitrary units), in excellent agreement with the value obtained from the FFT.

\begin{figure}
\includegraphics[width=\columnwidth]{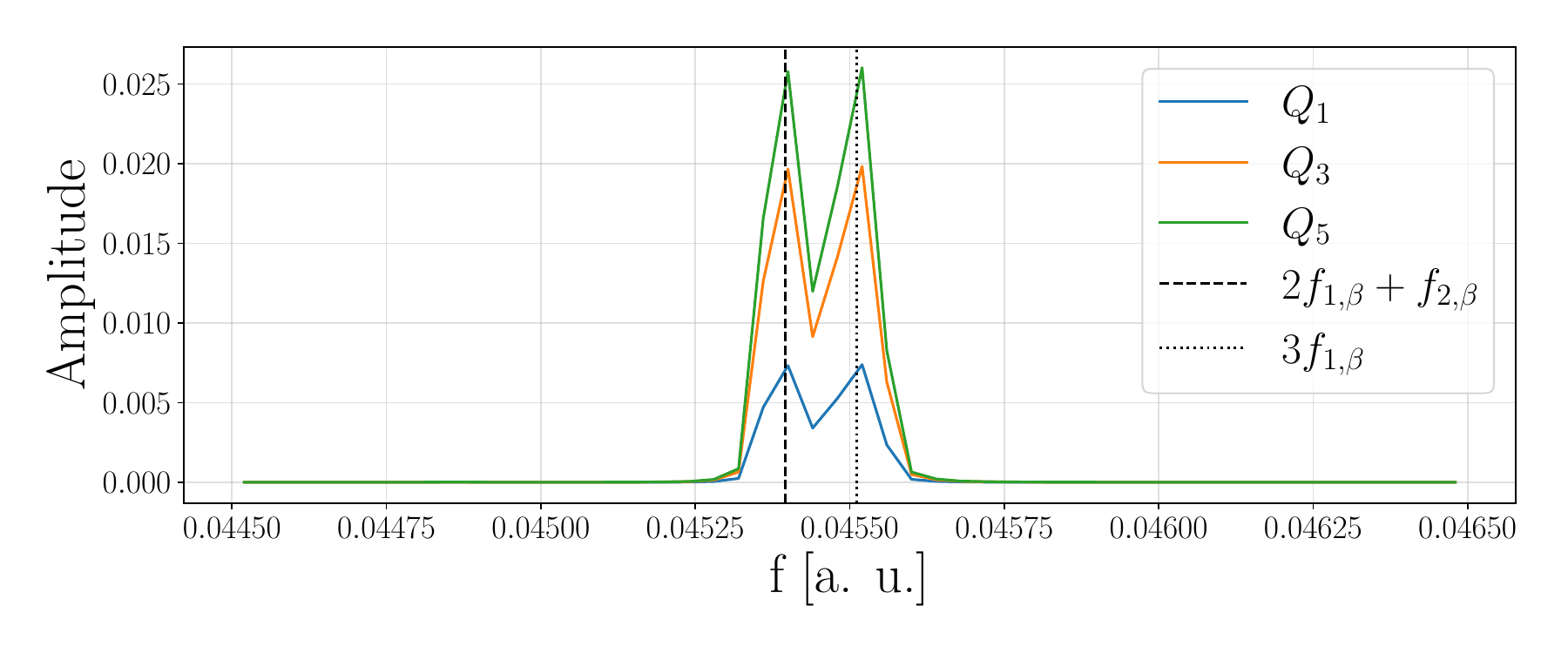}
\caption{FFT spectra of the normal mode coordinates $Q_1$, $Q_3$, and $Q_5$ for $\beta =0.1$ in the region of the third harmonic of frequency $f_{1, \beta} \approx 1.5170 \times 10^{-2}$. The spectra were computed using a Hann window to reduce spectral leakage.}   
\label{fig:fig9}      
\end{figure}

Regarding the second carrier frequency $\omega_{2,\beta}$, it is worth noting that it must satisfy the conditions $\omega_{2,\beta} \approx \omega_{1,\beta}$ and $\omega_{\rm rec,\beta}=\omega_{1,\beta}-\omega_{2,\beta}$, where $\omega_{\rm rec,\beta}$ denotes the recurrence (beat) frequency~\cite{Feynman1970Vol1}. In the present case, no spectral peak can be directly associated with $f_{2,\beta}$. Nevertheless, if such a second frequency exists, it can be inferred from the second relation above involving the recurrence frequency. This is indeed the case. We note that spectral doublets emerge at the harmonics of $f_{1,\beta}$ in the spectra of both $Q_i$ and $E_i$ ($i=1,3,5$). As an illustrative example, let us consider the doublet occurring at the third harmonic, $3f_{1,\beta}$, shown in Fig.~\ref{fig:fig9}. First, the frequency separation between the two peaks is found to be precisely the recurrence frequency, namely $\Delta f_\beta = f_{\rm rec,\beta}$. Second, the frequency of the lower sideband can be written as $2f_{1,\beta}+f_{2,\beta}$, from which it immediately follows that $f_{2,\beta}=f_{1,\beta}-f_{\rm rec,\beta}$. This is precisely the relation expected for beats arising from the superposition of two cosine waves with slightly different frequencies. Accordingly, we obtain $f_{2,\beta}\approx1.5055\times10^{-2}$ in arbitrary units.

\begin{table*}[t]
\centering
\caption{Frequencies $f$ and $\omega$, and the corresponding periods $T$ (all in arbitrary units), obtained from the Fourier analysis presented in the text. For the first mode ($k=1$, $\beta=0$, $N=32$), $\omega_1 = 2\sin(\pi/66)$. Numbers in parentheses denote powers of $10$.}
\label{tab:mode_frequencies}
\setlength{\tabcolsep}{16pt}

\begin{tabular}{cccc}
\hline
$\beta$ & $f$ & $\omega$ & $T$ \\
\hline
$\beta=0.0$
& $f_1 \approx 1.5146(-2)$
& $\omega_1 \approx 9.5164(-2)$
& $T_1 = 66.02$ \\

$\beta=0.1$
& $f_{1,\beta}\approx1.5170(-2)$
& $\omega_{1,\beta}\approx9.5317(-2)$
& $T_{1,\beta} \approx 65.92$ \\

$\beta=0.1$
& $f_{2,\beta}\approx1.5055(-2)$
& $\omega_{2,\beta}\approx9.4594(-2)$
&  $T_{2,\beta} \approx 66.42$ \\

$\beta=0.1$
& $f_{\rm rec,\beta}\approx1.15(-4)$
& $\omega_{\rm rec,\beta}\approx7.24(-4)$
&  $T_{\rm rec,\beta} \approx 86.80 (2)$ \\
\hline
\end{tabular}

\end{table*}

For completeness, we report the relevant numerical results of the Fourier analysis in Table~\ref{tab:mode_frequencies}. Furthermore, comparing $\omega_{1,\beta}$ and $\omega_{2,\beta}$ with the first-mode frequency $\omega_1$ of the corresponding harmonic chain with $N=32$ (see Table~\ref{tab:mode_frequencies}) shows that only very small frequency shifts are induced by the weak nonlinearity, namely approximately $0.16\%$ and $0.63\%$, respectively.

Overall, the Fourier analysis provides strong evidence that the dynamics of the FPUT $\beta$ model with $N=32$ in the recurrence regime can be described by the superposition of two cosine waves with independent  frequencies and different amplitudes. This finding is fully consistent with the intrinsic dimensionality estimate $m^{\ast}=2$ obtained using the data-driven approach based on a deep autoencoder model. 
Indeed, the DAE learns $z_{1, \beta}$ and $z_{2, \beta}$ that provide a compact and meaningful representation of the system's dynamics.
Consequently, these latent variables are naturally expected to be related to the angular coordinates $\phi_{1,\beta}$ and $\phi_{2,\beta}$ parameterizing the invariant torus, whose time evolution is governed by the two independent frequencies $\omega_{1,\beta}$ and $\omega_{2,\beta}$, respectively.

\section{Conclusion}\label{sec:conclusion}

In this work, we investigated the intrinsic dimensionality of high-dimensional trajectories in the FPUT chain using a deep learning approach. To this end, we employed a relatively simple deep autoencoder model to analyze the trajectory data. We find that, in the weakly nonlinear regime where energy recurrences occur ($\beta \lesssim 1$), the orbits lie on a small two-dimensional nonlinear manifold. This result is consistent with the previous analysis based on multi-chart flows~\cite{Yu2025}, which suggests that the underlying manifold possesses a Riemannian structure. It is further independently corroborated by the Fourier analysis presented here, which reveals two independent  frequencies, consistent with quasi-periodic motion on a two-dimensional invariant torus $\mathbb{T}^2$.

In contrast, a comparison with a linear approach based on principal component analysis and standard heuristics shows that it cannot capture the nonlinear structure of the data. As a result, PCA can only provide useful upper bounds for the intrinsic dimensionality. Furthermore, principal component analysis is unable to detect a crucial dynamical change, that is, the symmetry-breaking occurring at $\beta = 1.1$. In this case, the autoencoder predicts an increase in the intrinsic dimension by unity, which may be associated with significantly different geometric and topological features of the resulting three-dimensional nonlinear manifold.

In future work, it would be highly desirable to validate our findings using alternative manifold learning methods~\cite{Zeng_2024, Yu2025, zelong2025}. We note that the manifold hypothesis, namely, that high-dimensional data lie near or on low-dimensional manifolds~\cite{Fefferman2016,meila2024}, is strongly supported, in the present context, by the applicability of the KAM theorem in the weakly nonlinear regime of the $\beta$ model~\cite{lepri2023}.

\begin{acknowledgments}
\noindent The author thanks S\o{}ren Hauberg for advice on correctly training the deep autoencoder, Fabio Frohberg for helpful comments on the manuscript and for allowing the use of his JAX-based deep autoencoder script, and Alberto Maspero for explaining the Cauchy problem for the FPUT system as a first-order differential system. Most of the code execution for this project was performed using Google Colaboratory. The author is also grateful to Google for supporting this research through the Tensor Research Cloud (TRC) program and for providing access to Cloud Tensor Processing Units (TPUs), which accelerated the machine learning experiments. Finally, the identification of the spectral peaks was carried out with the assistance of the AI system Gemini (Google).

\end{acknowledgments}

\section*{Data Availability Statement}

The data that support the findings of this study are openly available in Zenodo~\cite{marchetti_2025_a, marchetti_2025_b}.

\appendix

\section{FPUT Initial Condition}\label{sec:fput_facts}

In the following, we recall how the ``half sine'' initial condition of FPUT chain is defined a time $t=0$~\cite{FORD1992}. Such a  condition corresponds to initially displace the coordinates $q_i$ ($i=1, \cdots, 32$) according to the formula~\cite{giordano2006} 
\begin{equation}\label{eq:initial_cond}
q_i\left(0\right) = A \sqrt{\frac{2}{N+1} }\sin\left( \frac{ik\pi}{N+1}\right) \, ,
\end{equation}
where the amplitude $A$ and wave number $k$ are set to $A=10$, and  $k=1$, respectively. Note that here $q_0= q_{N+1} = 0$, that is, the chain's ends are fixed. Additionally, it is assumed  that the variables $p_i \left(0\right) =0$ with $i =1, \cdots, 32$.  

Next, the $k$-th energy (Fourier) mode is defined as~\cite{giordano2006}
\begin{equation}\label{eq:energy_mode}
E_k = \frac{1}{2} \left[ \Dot{Q}_{k}^2  + \omega_{k}^{2}  Q_{k}^2\right]  \, ,
\end{equation}
where $\omega_{k} = 2 \sin\left(k\pi/2\left(N+1\right)\right)$ is the frequency,  $Q_k$ denotes the normal mode coordinate ($k=1, 2, \cdots, N$) which reads~\cite{giordano2006}
\begin{equation}\label{eq:variables}
Q_k = \sqrt{\frac{2}{N+1} } \sum_{j=0}^{N} q_j \sin\left( \frac{jk\pi}{N+1}\right) \, .
\end{equation}

In the present case, the large-wavelength (or equivalently low-frequency)  initial condition is equivalent to initially give the  energy $\mathcal{E}_1 \approx 0.45$ to first energy  mode $E_1$ only. 

In the weakly nonlinear regime ($\beta \lesssim 1$) the recurrence phenomenon occurs. As a result, the  energy is shared by three modes with odd wave numbers $k = 1, 3, 5$, while it cannot spread to modes with even wave numbers due to the parity conservation characterizing the $\beta$ model~\cite{pace2019}. This fact is illustrated in Fig.~\ref{fig:fig9} for $\beta = 1$.

\begin{figure}
   \includegraphics[width=\columnwidth]{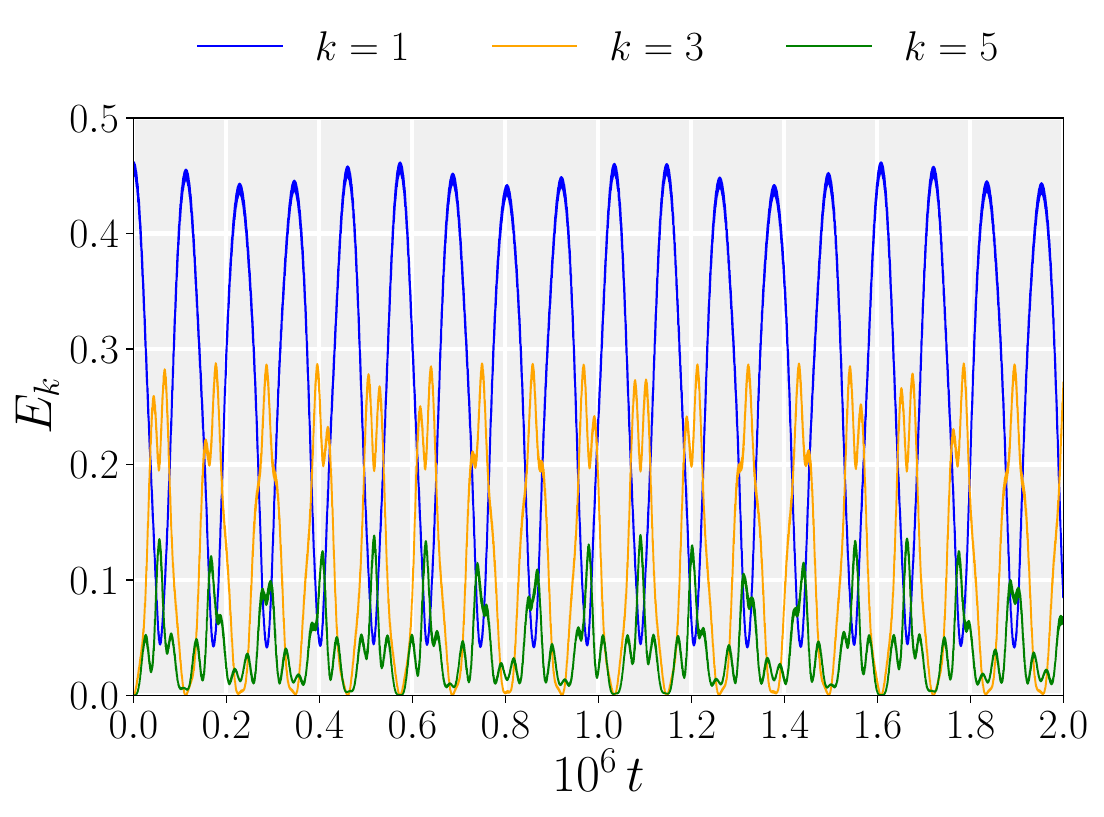}
\caption{Recurrence phenomenon. Time evolution of the energy $E_k$ of the Fourier modes with $k = 1, 3, 5$ in the FPUT $\beta$ model with $N = 32$ coupled oscillators ($\beta = 1$). The initial condition corresponds to assigning the energy $\mathcal{E}_1 \approx 0.45$  to the first mode $E_1$ only.}   
\label{fig:fig10}      
\end{figure}

Next, we show that the energy transfer to modes $E_2$ and $E_4$ is a direct consequence of the symmetry-breaking feature of the $\beta$ model, rather than an artifact of symplectic integration errors. To this end, we define the following energy fractions:
$\left(E_1 + E_3 + E_5\right)/\mathcal{E}_1,\quad
\left(E_2 + E_4\right)/\mathcal{E}_1$, and
$\left(\sum_{i=6}^{32} E_i\right)/\mathcal{E}_1$.

We then examine their time evolution during the system dynamics for
$\beta = 1.1$ (see Fig.~\ref{fig:fig10}). Toward the end of the simulation, the energy of the odd modes $E_1$, $E_3$, and $E_5$ (blue curve) decreases by approximately $6.55\%$, while the even modes $E_2$ and $E_4$ (brown curve) gain about $6.02\%$ of the initial energy $\mathcal{E}_1$. The difference between these values is therefore small, about $0.53\%$, confirming that an energy redistribution between these modes is taking place.

We also note that the energy of the remaining modes with $k \geq 6$ (grey curve) oscillates within the range $0\%-20\%$, indicating that no systematic energy accumulation occurs in these modes. We therefore conclude that the observed symmetry-breaking is not a fictitious effect arising from numerical integration errors.

\begin{figure}
\includegraphics[width=\columnwidth]{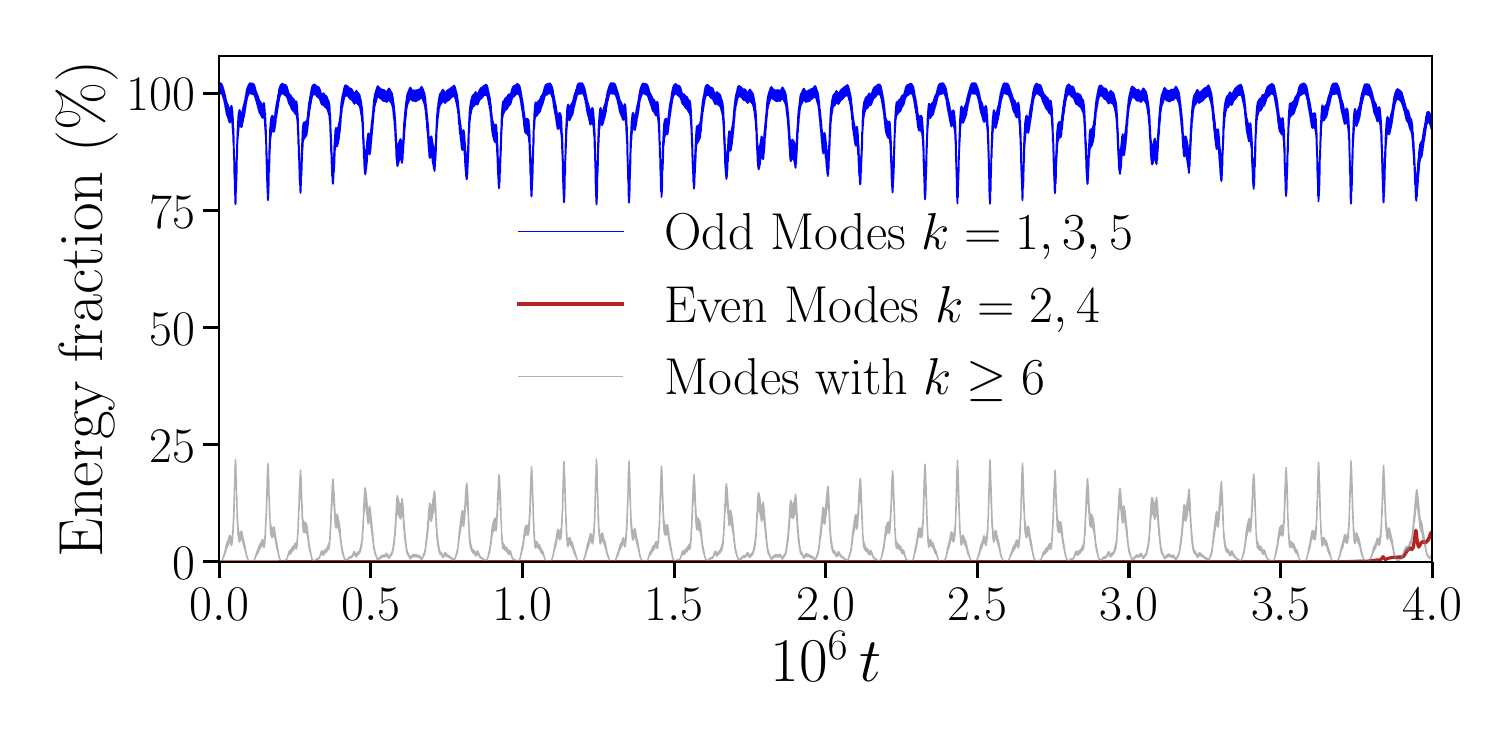}
\caption{ Time-evolution of the energy fractions (in percentage) $\left(E_1 + E_3 + E_5\right)/\mathcal{E}_1$ (odd modes, blue curve),  $\left(E_2 + E_4\right)/\mathcal{E}_1$ (even modes, brown color) and 
$\left(\sum_{i=6}^{32}E_i\right)/\mathcal{E}_1$ (modes with $k\geq 6$, grey curve), assuming $\beta=1.1$.}   
\label{fig:fig11}      
\end{figure}

\section{Principal Component Analysis}\label{sec:pca} 

Let  $X^{\ast}$ denote the $n_s \times n$  data matrix, after the data preprocessing, which make the variables  scale-free each with zero mean, then the sample covariance matrix (or equivalently correlation matrix) $S$ reads~\cite{Joliffe2016}
\begin{equation}\label{eq:covariance}
S =\frac{1}{n_s -1 }  X^{\ast^{T}} X^{\ast} \, .
\end{equation}

The eigenvalue $\lambda_i$  of $S$  correspond to the explained (or preserved) variance along the $i$-th principal component axis. In present work, the sorted eigenvalues  $\lambda_1 \geq \lambda_2 \geq \cdots \lambda_i \geq \cdots \lambda_n$  are efficiently computed  using the singular value decomposition~\cite{strang1993, stewart1993}. Accordingly to SVD,  $X^{\ast}=WLV^T$ where $W$ and $V$ are two orthogonal matrices, and $L$ is a diagonal matrix~\cite{Joliffe2016, Greenacre2022}.  As a result, it is found that~\cite{Joliffe2016}
\begin{equation}\label{eq:covariance1}
\lambda_i = \frac{1}{n_s -1} s_{i}^{2} \, ,
\end{equation}
where the singular values $s_i$  ($s_{1}^{2} \geq s_{2}^{2}  \geq \cdots \geq s_{n}^{2} \geq 0 $) are the diagonal entries of $L$, which we computed using scikit-learn library~\cite{pedregosa2011}.

From the projection perspective PCA orthogonally projects the original data points $\mathbf{x_j}$ onto projected data points $\mathbf{\tilde{x}_j}$ belonging to a suitable $m$-dimensional linear subspace (the principal subspace), by minimizing the average reconstruction error $J_m$ which reads~\cite{hastie2017}
\begin{equation}\label{eq:rec_error}
J_m = \frac{1}{n_s}\sum_{j= 1}^{n_s} \lVert \mathbf{x_j} - \mathbf{\tilde{x}_j} \rVert_2^{2} \, ,
\end{equation}
where $\lVert  \cdot \rVert_2$ stands for the Euclidean norm.

Accordingly, within PCA,  Eq.~\ref{eq:rec_error} can be rewritten in terms of covariance matrix's eigenvalues $\lambda_i$ as~\cite{Deisenroth2020}
\begin{equation}\label{eq:rec_error_1}
J_m = \sum_{l= m + 1}^{n} \lambda_l \, .
\end{equation}

It is worth noting here that the eigenvectors relative to the eigenvalues $\lambda_l$ ($ m + 1 \leq l  \leq n$) in Eq.~\ref{eq:rec_error_1} constitute the basis of the orthogonal complement of the principal subspace. 

Finally, it is worth noting that within PCA, the term PC score vector (or simply scores) replaces the code vector which sits at the AE's bottleneck layer~\cite{Joliffe2016}.

\subsection{Heuristics for Determining ID in PCA}\label{sec:heuristics}

Determing the intrinsic dimensionality corresponds to find the optimal number of PCs to retain. To this end, there exist various methods. To our best knowledge, the state-of-art method is represented by the Gavish-Donoho optimal hard threshold, which rests on the Marchenko-Pastur distribution~\cite{Gavish2014, marcenko1967}. However, this cannot applied here because the aspect ratio $n/n_s \approx 1.6 \times 10^{-5}$ of $\tilde{X}$ is too small~\cite{marchetti2025}.

Jolliffe modified  the Kaiser criterion (also known as the Kaiser-Gutman rule) in factor analysis, making it suitable for PCA~\cite{Kaiser1960, jolliffe2002pca}. According to Jolliffe's  ansatz,  the number of principal components which capture most of the variance, corresponds to those PCs for which $\lambda_i \geq 0.7$.

An alternative way to obtain the ID from  PCA's eigenvalues is computing the Participation Ratio $D_{\rm PR}$, which estimates the number of dimensions along which the data is distributed~\cite{Kramer1993, RECANATESI2022}. Accordingly, this quantity reads~\cite{RECANATESI2022}
\begin{equation}\label{eq:participation}
D_{\rm PR} = \frac{\left( \sum\limits_{i=1}^{n} \lambda_i \right)^2}{\sum\limits_{i=1}^{n} \lambda_i^2} \, .
\end{equation}

Eq.~\ref{eq:participation} can be conventiently expressed in terms of the traces of $S$ (Eq.~\ref{eq:covariance}) and $S^{2}$, respectively. In such a case, one finds $D_{\rm PR} = \left( \operatorname{Tr}(S) \right)^2/\operatorname{Tr}(S^2)$. Since $D_{\rm PR}$  is as an estimate of the intrinsic dimensionality of the data, an inherently integer-valued quantity, we round it to the nearest integer.

\section{DAE Reconstruction Errors for $\beta=1.8$}\label{sec:TPUs}

In Fig.~\ref{fig:fig11}, the DAE and PCA reconstruction error curves as functions of the latent dimension $m$ are shown for $\beta = 1.8$. In this case, most computations were performed using Google TPUs; accordingly, the only modification to the autoencoder architecture (ReLU, $N_h=5$, Adam) was an increase in the batch size from $256$ to $512$ in order to fully exploit the TPUs. However, this change in the autoencoder's hyperparameters does not affect the overall final result.

Although the curves suggest that the intrinsic dimensionality increases substantially for large $\beta$, no clear elbow points are visible, making their identification highly questionable. Accordingly, by applying KA to the DAE reconstruction curve, and PR, KC, and KA to the PCA reconstruction curve, we obtain the estimates $m^{\ast} = 6$ for DAE, and $m^{\ast} = 6$, $m^{\ast} = 17$, and $m^{\ast} = 21$ for PCA, respectively.

Assuming, as discussed above, that the PR heuristic provides the most reliable estimate of the intrinsic dimensionality in the linear approach, we conclude that both methods predict the same intrinsic dimension, $m^{\ast} = 6$, for data with $\beta = 1.8$. Furthermore, DAE aligns with PCA, confirming that the intrinsic dimensionality increases with stronger nonlinearity, as previously reported in Ref.~\cite{marchetti2025}.

\begin{figure}
   \includegraphics[width=\columnwidth]{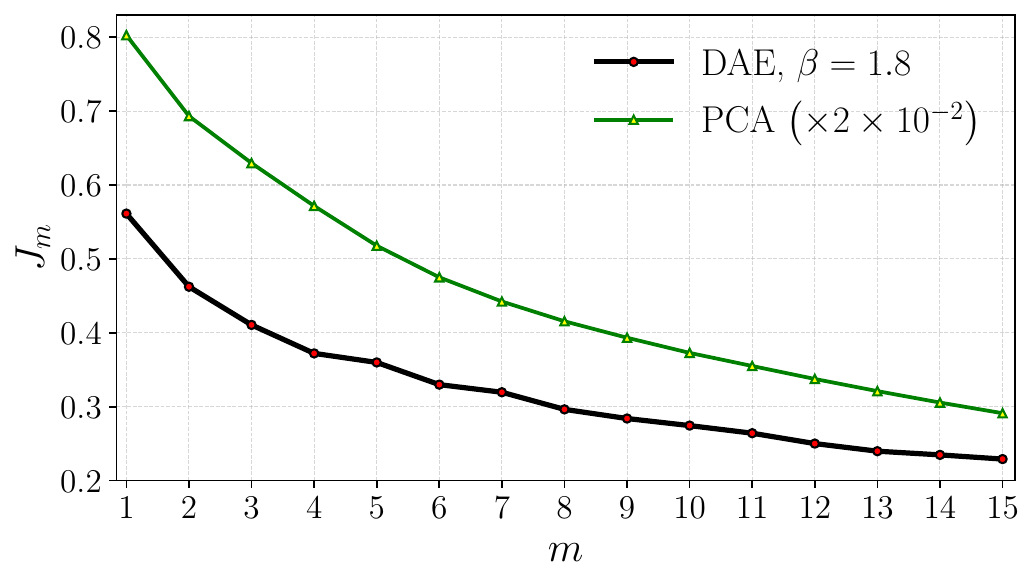}
\caption{DAE (ReLU, $N_h=5$, Adam) reconstruction error curve (circle) and PCA reconstruction error curve (triangle), rescaled by a factor of $50$, as functions of latent dimension $m$ for trajectory data corresponding to $\beta= 1.8$.}   
\label{fig:fig12}      
\end{figure}

\section{Fourier Analysis of Signals $Q_k$ and $E_k$ ($k=1, 3, 5$, $\beta=0.1$)}\label{sec: spectra}

In the following, we briefly report the frequency analysis of the normal mode coordinates $Q_1$, $Q_3$, and $Q_5$, together with their respective energies $E_1$, $E_3$, and $E_5$, obtained using the Fast Fourier Transform  with a Hann window to reduce spectral leakage. The FFT was implemented using the SciPy Python library.

Accordingly, the FFT was applied to datasets consisting of $500,000$ values of $Q_k$ and $E_k$ for each mode index $k=1,3,5$, assuming $\beta=0.1$. As a result, the frequencies can be resolved with a resolution of approximately $4\times10^{-5}$ in arbitrary units.

\begin{figure}[t]
    \centering

    \includegraphics[width=\columnwidth]{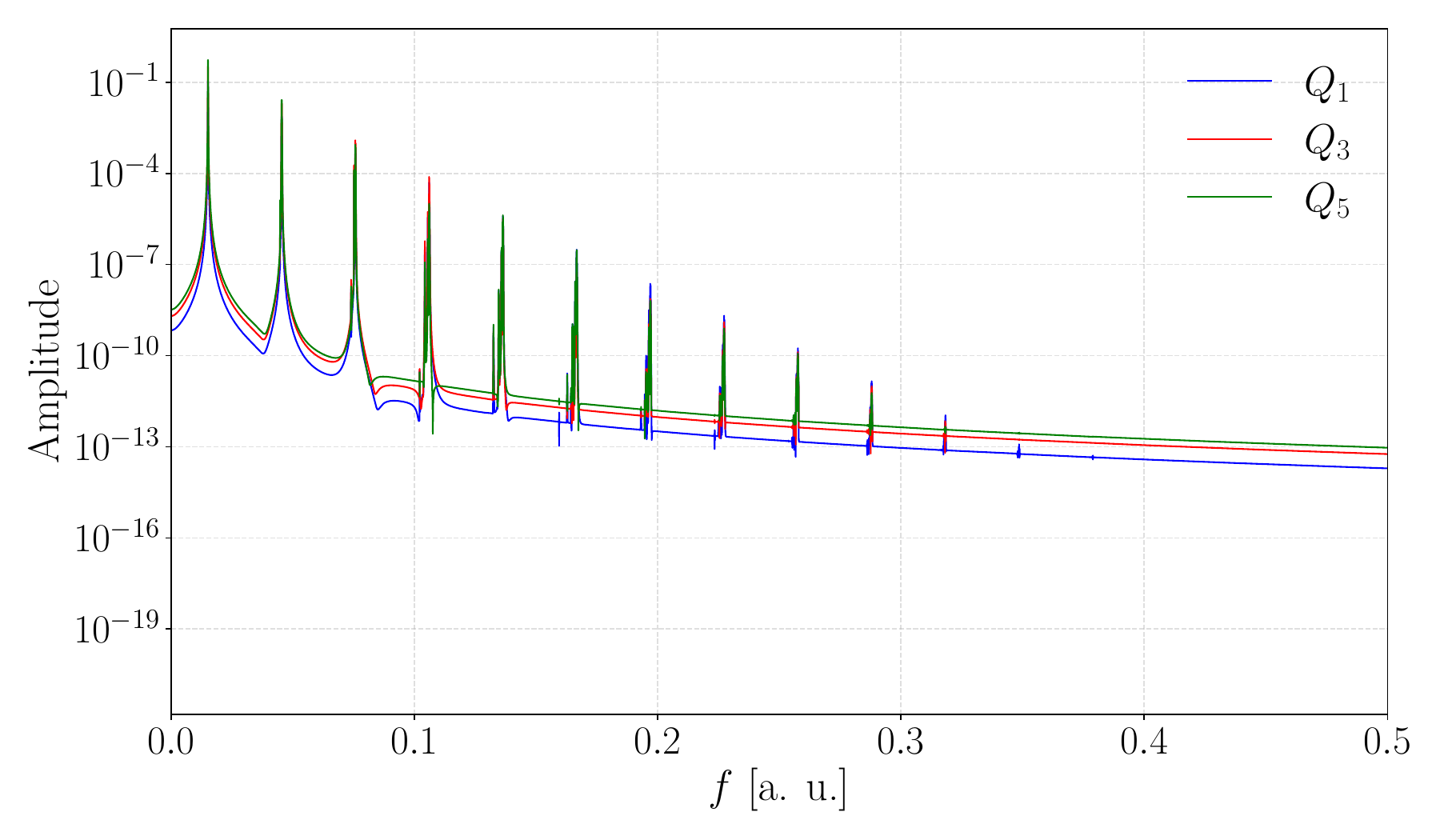}

    \vspace{0.5em}

    \includegraphics[width=\columnwidth]{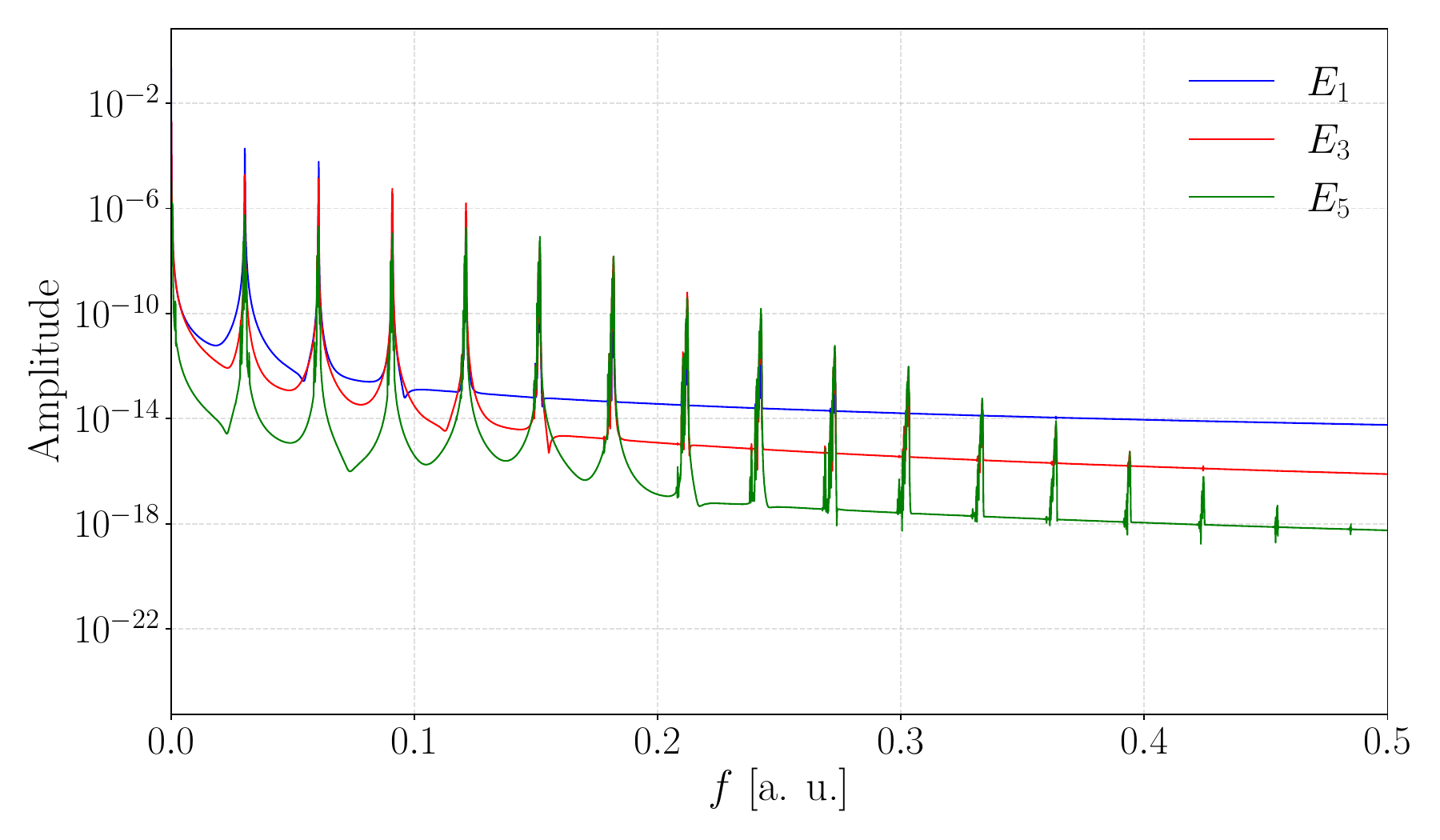}

    \caption{FFT spectra of the normal mode coordinates $Q_1$, $Q_3$, and $Q_5$ (top panel) and of the corresponding  energies $E_1$, $E_3$, and $E_5$ (bottom panel). The spectra were computed using a Hann window to reduce spectral leakage.}
    \label{fig:fig13}
\end{figure}

The spectra of $Q_1$, $Q_3$, and $Q_5$ show significant peaks at the frequency $f_{1,\beta}\approx1.5170\times10^{-2}$ a.u. and at its odd harmonics, i.e., $f = (2 n + 1) f_{\rm 1, \, \beta}$ where $n=1, 2, \cdots$ (see the top panel of Fig.~\ref{fig:fig13}). Note that the estimated value of $f_{1,\beta}$ is very close to the theoretical frequency $f_1$ of the corresponding linear system reported in Table~\ref{tab:mode_frequencies}. In fact, their small difference is approximately $0.16\%$

Next, the bottom panel of Fig.~\ref{fig:fig13} shows the spectra of the corresponding energies $E_1$, $E_3$, and $E_5$ as functions of the frequency $f$. It is found that the dominant peaks appear at the even harmonics of $f_{1,\beta}$, i.e., $f = 2 n f_{1, \beta}$ where $n=1, 2, \cdots$, confirming that this should be one of the two carrier frequencies contributing to the beats. Note that the appearance of the even harmonics of $f_{1,\beta}$ in energy spectra is a direct consequence of the quadratic dependence of the energy on the normal coordinate, namely $E_i\propto Q_i^2$.

The same spectra also display  peaks at $f_{\rm rec, \beta} \approx 1.15 \times 10^{-4}$ although they are not visible here because of their small amplitude. For instance, the amplitudes of $E_1$ and $E_3$ are $\approx 0.002$. This is the expected beat (or recurrence) frequency responsible for the slow modulation of the signal amplitude. It can also be estimated directly from the separation between successive maxima of the signals $Q_k$ and $E_k$, yielding a recurrence period
$T_{\rm rec,\beta}\approx8680$ a.u., which satisfies the expected relation $f_{\rm rec,\beta}=1/T_{\rm rec,\beta}$.

Finally, the second carrier frequency $f_{\rm 2,\beta}$ cannot be measured directly, but inferred by looking for spectral peaks whose frequencies satisfies the relationship $f_{\rm 2, \beta} = f_{\rm 1,\beta} -  f_{\rm rec, \beta}$. In this regard, we refer the reader to the discussion in Sec~\ref{sec:discussion}.

\bibliography{aipsamp}
\end{document}